\documentclass[twocolumn,showpacs,preprintnumbers,amsmath,amssymb]{revtex4}

\usepackage{graphicx}
\usepackage{dcolumn}
\usepackage{bm}

\begin{document}


\title{Density Matrix Renormalization Group and the Nuclear Shell Model}

\author{S. Pittel}

\affiliation{Bartol Research Institute, University of Delaware,
Newark, Delaware 19716, USA}

\author{N. Sandulescu}

\affiliation{Institute of Physics and Nuclear Engineering, 76900
Bucharest, Romania}

\affiliation{Service de Physique Nucleaire, CEA-Saclay,
  F91191 Gif sur Yvette, France}

\date{\today}

\begin{abstract}
We describe the use of the Density Matrix Renormalization Group
method as a means of approximately solving large-scale nuclear
shell-model problems. We focus on an angular-momentum-conserving
variant of the method and report test results for the nucleus
$^{48}Cr$. The calculation is able to reproduce both the ground
state energy and the energy of the first excited state, by
diagonalizing matrices much smaller than those of the full shell
model.
\end{abstract}

 \maketitle

\section{Introduction}

In the traditional nuclear shell model \cite{Heyde}, the
low-energy structure of a given nucleus is described by assuming
an inert doubly--magic core and then diagonalizing the effective
nuclear hamiltonian within an active  valence space consisting of
at most a few major shells. Despite the enormous truncation
inherent in this approach, the shell-model method as just
described can still only be applied in very limited nuclear
regimes, namely for those nuclei with a sufficiently small number
of active nucleons or a relatively low degeneracy of the valence
shells that are retained. For heavier nuclei or nuclei farther
from closed shells, one must truncate further to reduce the number
of shell-model configurations to a manageable size.

An attractive truncation possibility is provided by the Density
Matrix Renormalization Group (DMRG), a method initially developed
for low-dimensional quantum lattices \cite{White}. Subsequently,
the method was extended to finite Fermi systems and applied to
ultrasmall superconducting grains \cite{Duke1}, to problems in
quantum chemistry \cite{QC} and to two-dimensional electrons in
strong magnetic fields \cite{2D}. The great success of these
applications suggests that it could also prove useful for
obtaining accurate approximate solutions to the nuclear shell
model, in cases where exact diagonalization is not feasible.

The DMRG method involves a systematic inclusion of the degrees of
freedom of the problem. When treating quantum lattices, real-space
sites are added iteratively. In finite Fermi systems, these sites
are replaced by single-particle levels. At each stage, the system
[referred to as a {\em block}] is enlarged to include the
additional site or level. This enlarged block is then coupled to
the rest of the system (the {\em medium}) giving rise to the {\em
superblock}. For a given eigenstate of the superblock (often the
ground state) or perhaps for a group of important eigenstates, the
reduced density matrix of the enlarged block in the presence of
the medium is constructed and diagonalized and those states with
the largest eigenvalues are retained. This method of truncation is
guaranteed to be optimal in the sense that it maximizes the
overlap of the approximate (truncated) wave function with the
targeted superblock wave function prior to truncation.

The earliest application of the DMRG method in the context of
nuclear structure was based on the use of a particle-hole variant
of the method \cite{Sevi}. In this approach, the group of
single-particle levels was divided into four subgroups, neutron
particle, neutron hole, proton particle and proton hole. The
results of those calculations were not terribly encouraging,
however, as they required that a substantial fraction of the full
shell model space be retained for a reasonable reproduction of the
results from exact diagonalization.

More recently a more traditional DMRG study was reported that did
not adopt the particle-hole strategy \cite{Papenbrock}.  The
method worked extremely well for the nucleus $^{28}Si$, giving an
accurate reproduction of the exact shell-model results with a
significantly reduced basis. In contrast, the results for
$^{56}Ni$ were not nearly as good, as the method seemed to be
converging to a solution energetically still far from the exact
ground state.

One of the limitations of the above studies is that they did not
preserve symmetries throughout the iterative inclusion of
single-particle levels. Since the density matrix procedure
involves a truncation at each of the iterative stages, there is a
potential to lose these symmetries and the associated
correlations. In such cases, the iterative procedure must
incorporate not only dynamical correlations but also the
kinematical correlations associated with symmetry restoration.

This point has been recognized for some time in DMRG applications
outside the nuclear domain. McCulloch and Gul\'{a}csi
\cite{McCulloch} discussed how to reformulate the DMRG to maintain
symmetries throughout. They referred to this as the non-Abelian
DMRG method and showed that it could produce more accurate results
than the traditional Abelian method while using smaller spaces.

In this work, we adopt a strategy whereby angular momentum is
preserved throughout the iterative DMRG process. We refer to this
as the JDMRG method. It was briefly described as part of a recent
review \cite{Review}. Subsequently, the method was applied in
nuclear physics for the first time in the context of the Gamow
Shell Model \cite{PloPlo}.

An outline of the paper is as follows. In Sec. II, we provide a
brief overview of the traditional Abelian DMRG method and then in
Sec. III describe the modifications required to maintain
rotational invariance as an exact symmetry. In Sec. IV, we briefly
review the test problem we treat and describe in some detail our
implementation of the JDMRG method for this problem. Section V
describes our results. Section VI summarizes the principal
conclusions reached in this study and outlines directions for
future work.

\section{Overview of the DMRG method}

We begin by describing the traditional Abelian DMRG method, as has
been used in almost all applications to date.  We focus on finite
Fermi systems, for which the degrees of freedom being
systematically incorporated by the method are single-particle
levels.

Assume we have treated a set of $L$ levels and that the total
number of states we have kept to describe that {\em block} is $m$.

We now add the next level, the $L+1^{st}$, which admits $s$
states. In the traditional DMRG approach, the states we build in
the enlarged block are simply products of those in the original
block and in the extra level. Thus, when we enlarge the block to
include the $L+1^{st}$ level, it results in a new block with  $m
\times s$ states.

Renormalization Group methods, whether Wilson's original numerical
algorithm or White's Density Matrix version, then implement a
truncation of the states of the enlarged block to the same number
$m$ as before the block enlargement.

In Wilson's numerical RG procedure, the truncation involves
diagonalizing the hamiltonian in the $m \times s$-dimensional
space of the enlarged block and retaining its lowest $m$
eigenstates.

In White's DMRG approach, the truncation is implemented through a
very different procedure, schematically illustrated in figure 1.
Consider the enlarged block $B^{\prime}$ (the block $B$ plus the
added level)  in the presence of a medium $M$ that approximates
the rest of the system. White's truncation is carried out based on
the importance of the block states in a selected set of target
states of the full {\em superblock}, {\em i.e.,} the eigenstates
of $B^{\prime}$ coupled to $M$.

\begin{figure}
  \includegraphics[height=.2\textheight]{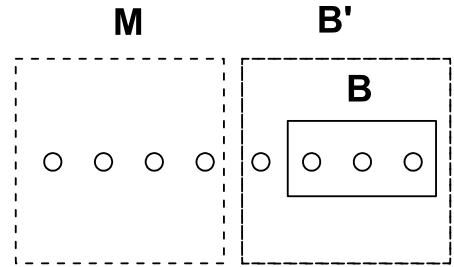}
  \caption{Schematic illustration of the DMRG growth procedure,
  in which a block $B$
  is enlarged to $B^{\prime}$ in the presence of a medium $M$.}
\end{figure}

Assume here that we target only the ground state of the superblock
in the truncation,

\begin{equation}
| \Psi_{gs} \rangle = \sum_{i=1,~m \times s}~\sum_{j=1,~t}
~\Psi_{ij} ~|i \rangle_B ~| j \rangle_M ~,
\end{equation}
obtained by constructing and then diagonalizing the superblock
hamiltonian. Here $|i \rangle_B$ refers to a state in the block
and $|j \rangle_M$ to a state in the medium. Furthermore,  $t$
denotes the number of states of the medium. If we construct the
ground-state density matrix for the enlarged block,
\begin{equation}
\rho^B_{ii'} ~=~ \sum_{j=1,~t} \Psi^*_{ij} \Psi_{i'j}~,
\end{equation}
diagonalize it, and truncate to the $m$ states with the largest
density-matrix eigenvalues, we are guaranteed to achieve the
optimal approximation to the superblock ground state.

To target a group of states, we would construct a mixed density
matrix containing information on the block content of all of them.
For example, if we wished to target both the ground state and the
first excited state, weighting them equally, we would construct
\begin{equation}
\rho^B_{ii'} ~=~ \frac{1}{2}~\left(\sum_{j=1,~t} \Psi^*_{ij}
\Psi_{i'j}~+~\sum_{j=1,~t} \Phi^*_{ij} \Phi_{i'j} \right),
\end{equation}
where $\Psi$ as above refers to the ground state wave function and
$\Phi$ correspondingly to the wave function of the first excited
state.

Truncation to the most important $m$ states in the enlarged block
is accompanied by the renormalization of all operators for use in
the truncated space. If the eigenstates $|\alpha \rangle$
associated with the $m$ lowest density matrix eigenvalues are
\begin{equation}
|\alpha \rangle = \sum_{i=1,m \times s} c^{\alpha}_i |i \rangle
~~~,~~\alpha=1,...,m~,
\end{equation}
then the matrix elements of any operator $O$ must be renormalized
according to
\begin{equation}
\langle \alpha | O | \beta \rangle  = \sum_{i,j} c^{\alpha *}_i
c^{\beta}_j ~ \langle i | O | j \rangle ~.
\end{equation}

A key step in the growth procedure is to calculate at each step
the matrix elements of all necessary sub-operators of the
hamiltonian and to store them. This includes all one- and
two-point operators,

\begin{equation}
a^{\dagger}_i, ~ a^{\dagger}_ia_j, ~ a^{\dagger}_ia^{\dagger}_j,
~+h.c.,
\end{equation}
and all three- and four-point operators required to build the
hamiltonian matrix,

\begin{displaymath}
\hat{O}^1_k = \frac{1}{4}\sum_{ijl} V_{ijkl} a
^{\dagger}_ia^{\dagger}_ja_l ~+h.c. ~,
\end{displaymath}
\begin{equation}
\hat{O}^2=\frac{1}{4}\sum_{ijkl}V_{ijkl}~a^{\dagger}_ia^{\dagger}_ja_la_k
~,
\end{equation}
where $V_{ijkl}=\langle ij|V|kl \rangle$. Having this information
for the block and the additional level enables it to be calculated
for the enlarged block.

As an example, consider the one-body operator
$a^{\dagger}_{\alpha}a_{\beta}$. Its matrix elements in the
enlarged product space $|i,j \rangle = |i \rangle |j \rangle$,
where $|i \rangle$ belongs to block $B_1$ and $|j \rangle$ belongs
to $B_2$, are given by

\begin{eqnarray}
\langle i, j | a^{\dagger}_{\alpha} a_{\beta} | k,l \rangle
&=&\langle i | a^{\dagger}_{\alpha} a_{\beta} | k \rangle
~\delta_{jl} \nonumber \\
&+&\langle j | a^{\dagger}_{\alpha} a_{\beta} | l \rangle
~\delta_{ik} \nonumber \\
&+& (-)^{n_k} \langle i | a^{\dagger}_{\alpha} | k \rangle ~
\langle j | a_{\beta} | l \rangle \nonumber \\
&-& (-)^{n_k} \langle i | a_{\beta} | k \rangle ~ \langle j |
a^{\dagger}_{\alpha} | l \rangle ~,  \label{enlarge}
\end{eqnarray}
where $n_i$ is the number of particles in state $|i \rangle$. Note
that the matrix elements of a two-point operator in the product
space is expressed in terms of matrix elements of one- and
two-point operators in the spaces of the two component blocks.

We now have the basic tools in hand to discuss how the traditional
DMRG procedure is implemented. It involves a series of steps. The
first is to choose an active space and a hamiltonian to act in
that space. Next we choose an order in which the single-particle
levels are included. Typically, this is done by organizing the
levels in a chain, as schematically illustrated in figure 2.

\begin{figure}
  \includegraphics[height=.15\textheight]{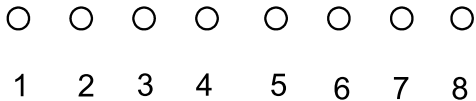}
  \caption{Schematic illustration of the ordering of levels (or sites) in a chain to be
  systematically included in the DMRG growth procedure.
}
\end{figure}

The following step is usually referred to as the warmup phase.
This step involves making an initial guess for the optimal
structure in each size block for a given choice of the number of
retained states $m$. For the eight levels shown in figure 2, we
would make a first guess at the structure of the blocks $B(1,2)$,
$B(1,3)$, $B(1,4)$, $B(1,5)$, $B(1,6)$, $B(1,7)$ and $B(1,8)$.
Having a guess for the optimal $m$ states in each block, we can
calculate (and store) all the necessary matrix elements in those
truncated blocks. The warmup guess can be implemented in a variety
of ways, depending on the specific problem.

The next step, and indeed the heart of the DMRG method, is the
sweep phase, which we now describe by focusing again on the chain
of eight levels in figure 2.

We begin the sweep phase by growing level $8$ into a two-level
block $B(7,8)$. To determine the optimal $m$ states in that
enlarged block, we immerse it in a medium consisting of the block
$B(1,6)$ from the warmup phase. We diagonalize the superblock
hamiltonian for the full system, then calculate the reduced
density matrix associated with the enlarged block, and renormalize
all operators to act in the truncated space associated with the
$m$ largest eigenvalues.

We then continue the process, enlarging to $B(6,8)$, immersing it
in a medium consisting of $B(1,5)$  and then truncating and
renormalizing $B(6,8)$ based on its reduced density matrix
eigenvalues.

This process is continued until all size blocks have been treated.

At this point, we simply reverse the direction of the sweep, now
growing the system from left to right, always treating the
enlarged block in a medium consisting of the rest of the levels as
obtained in the previous sweep. This process is continued until
acceptably small changes from one sweep to the next are achieved.
At this point, we have hopefully arrived at an optimal block
structure and an optimal description of the system for the assumed
value of $m$.

The calculation is then carried out as a function of $m$, until
acceptably small changes with increasing values are achieved. At
this point, we hopefully have a good approximation to the exact
solution of the problem. Obviously, the solution is especially
useful if this good approximation can be achieved while retaining
only a small fraction of the full Hilbert space.

\section{The JDMRG method}

We now describe the modifications to the traditional DMRG
algorithm required to preserve rotational symmetry throughout.

As noted earlier, the DMRG systematically grows the system by
adding additional degrees of freedom. This is most conveniently
accomplished by adding doubly-degenerate levels, either
Nilsson-like levels or the $+m$ and $-m$ levels of an $nlj$
multiplet. A great advantage of this is that every level looks
very similar to every other, greatly simplifying the algorithm by
which they are computationally included.

In the JDMRG, it is essential that throughout the procedure the
states have definite angular momentum. To accomplish this, it is
critical that we add at each stage states of definite angular
momentum. This is most readily accomplished by always adding a
complete shell, rather than a partial shell or a non-spherical
single-particle level.

The goal of a calculation is to construct and diagonalize the
hamiltonian in the the space in which all levels have been
included. As noted earlier, the traditional DMRG works in a
product (or m-scheme) representation. In this case, we need to
build the matrix elements of all hamiltonian sub-operators
directly in the m-scheme, as this is what is needed to build the
matrix elements in the enlarged system and ultimately the matrix
elements of the superblock hamiltonian. For each size block those
m-scheme matrix elements must then be stored, for use in the
following sweep.

In an angular-momentum-conserving approach we must instead
calculate and store throughout the iterative process the {\em
reduced matrix elements} of all relevant sub-operators of the
hamiltonian, namely

\[
a^{\dagger}_i, ~ [a^{\dagger}_i\tilde{a}_j]^K, ~
[a^{\dagger}_ia^{\dagger}_j]^K, \]
\begin{eqnarray}\hat{O}^1_l=-\sum_{ijkK} && \sqrt{(1+\delta_{ij})(1+\delta_{kl})}~\hat{K}~
V^K_{ijkl} \nonumber\\
&\times
&~\left([a^{\dagger}_ia^{\dagger}_j]^K\tilde{a}_k\right)^l,
\nonumber \\ \hat{O}^2=-\sum_{ijklK}
&&\sqrt{(1+\delta_{ij})(1+\delta_{kl})}~\hat{K}~
V^K_{ijkl} \nonumber \\
&&\times\left([a^{\dagger}_ia^{\dagger}_j]^K
~[\tilde{a}_k\tilde{a}_l]^K \right)^0~,
\end{eqnarray}
and the relevant hermitean conjugates. Here,
$\hat{K}=\sqrt{2K+1}$, $\tilde{a}_{jm}=(-)^{j-m}a_{j-m}$ and
\[
V^K_{ijkl}=\langle ij(K)|V|kl(K)\rangle.
\]

As an illustration of how this proceeds, consider the reduced
matrix elements of the coupled one-body operator
$[a^{\dagger}_i\widetilde{a}_j]^K$ in an enlarged space obtained
by coupling blocks $B_1$ and $B_2$. Denoting the coupled states as
$\left. |\alpha K,\beta L(J) \right \rangle$, where $ \left.
|\alpha K \right\rangle$ refers to the state in $B_1$, $\left.
|\beta L \right\rangle$ to the state in $B_2$, and $J$ to the
coupled angular momentum, the matrix elements are given by

\begin{eqnarray}
&&\langle \alpha' K',\beta' L'(J')||
[a^{\dagger}_i\widetilde{a}_j]^{\lambda} ||\alpha K,\beta L(J)
\rangle~= \nonumber \\
&&~~~\delta _{\beta \beta ^{\prime }}\delta _{LL^{\prime }}\left(
-\right) ^{K^{\prime }+L^{\prime }+J+\lambda }~\widehat{J}\widehat{J%
}^{\prime }\left\{
\begin{array}{ccc}
K^{\prime } & J^{\prime } & L^{\prime} \\
J & K & \lambda
\end{array}
\right\} \nonumber \\
&&~~~\times \left\langle \alpha ^{\prime
} K^{\prime }\right\| \left( a _{i}^{\dagger }\widetilde{a}%
_{j}\right) ^{\lambda }\left\| \alpha K\right\rangle + \nonumber
\\
&&~~~+\delta _{\alpha \alpha ^{\prime }}\delta _{KK^{\prime
}}\left( -\right) ^{K^{\prime }+L+J^{\prime }+\lambda }~\widehat{J}\widehat{J%
}^{\prime }\left\{
\begin{array}{ccc}
L^{\prime } & J^{\prime } & K^{\prime } \\
J & L & \lambda
\end{array}
\right\} \nonumber \\
&&~~~\times \left\langle \beta ^{\prime } L^{\prime }\right\|
\left( a_{i}^{\dagger }\widetilde{a }_{j}\right) ^{\lambda
}\left\| \beta
L\right\rangle + \nonumber \\
&& ~~~+\left( -\right)^{n_{\alpha
K}}~\widehat{J}\widehat{J}^{\prime }\widehat{\lambda }\left\{
\begin{array}{ccc}
K^{\prime } & L^{\prime } & J^{\prime } \\
K & L & J \\
i & j & \lambda
\end{array}
\right\}
\nonumber \\
 &&~~~\times \left\langle \alpha ^{\prime }
K^{\prime }\right\| a _{i}^{\dagger }\left\| \alpha
 K\right\rangle \times \left\langle \beta ^{\prime
} L^{\prime }\right\| \widetilde{a}_{j}\left\| \beta
L\right\rangle + \nonumber \\
&&~~~+\left( -\right) ^{i-j+\lambda }~\left( -\right)^{n_{\alpha K}}~\widehat{J}\widehat{J}^{\prime }\widehat{%
\lambda }\left\{
\begin{array}{ccc}
K^{\prime } & L^{\prime } & J^{\prime } \\
K & L & J \\
j & i & \lambda
\end{array}
\right\} \nonumber \\
&&~~~\times \left\langle \alpha ^{\prime } K^{\prime }\right\|
\widetilde{a}_{j}\left\| \alpha K\right\rangle \times \left\langle
\beta ^{\prime } L^{\prime }\right\| a _{i}^{\dagger }\left\|
\beta L\right\rangle ~,
\end{eqnarray}
where $n_{\alpha K}$ is the number of particles in state $|\alpha
K \rangle$ . Knowing the reduced matrix elements of the one- and
two-point operators $a^{\dagger}_i, ~
[a^{\dagger}_i\tilde{a}_j]^K, ~ [a^{\dagger}_ia^{\dagger}_j]^K$ in
the two component blocks enables us to determine the reduced
matrix element of interest for the enlarged block.

Otherwise, the JDMRG involves the same set of basic steps as the
traditional DMRG.

\section{The test model}

The test results we report in this paper are for the nucleus
$^{48}Cr$, treated as four valence neutrons and four valence
protons outside doubly-magic $^{40}Ca$. We restrict both the
valence neutrons and the valence protons to the orbits of the
$1f2p$ shell. Furthermore, we use in these calculations the shell
model hamiltonian KB3, for which exact results are available. The
size of the full space for $^{48}Cr$ is 1,963,461 states. Of
these, 41,355 are $0^+$ states, 182,421 are $2^+$ states, etc.

\subsection{Some specifics of our implementation for this test problem}

In the calculations reported here, we separate neutron from proton
orbitals, rather than putting them in a single chain. As we will
soon see, this leads to a three-block strategy for implementing
the JDMRG truncation.

With that as background, we now describe how we implement the two
stages of the analysis, the warmup stage and the sweeping stage,
for the $^{48}Cr$ test problem we have considered.

\subsubsection{Warmup}

As noted earlier, the initial warmup stage of the DMRG procedure
involves choosing a first approximation to the optimal truncated
structure of each group of single-particle levels. Because of our
separation of neutrons from protons, the groups of relevance are
groups of neutron orbits and separately groups of proton orbits.
Labelling the four orbits as $1 \rightarrow 4$ (note that we have
not yet defined the order of the four active orbits), we need
information on the following blocks: $B(1,2)$, $B(1,3)$, $B(1,4)$.
While we need them for both neutrons and protons, the symmetry of
the problem allows us to choose them the same for both.

Our procedure is to gradually build the identical orbit block, by
systematically adding orbit $i+1$ to the block $B(1,i)$. At each
stage of the warmup growth, we construct a {\em superblock} by
coupling the enlarged neutron block to the corresponding enlarged
proton block ({\em i.e.,} the block with the same set of orbits).
We construct the superblock for N=Z, from $0 \rightarrow 4$, and
for $J^{\pi}=0^+$. Targeting the ground state for each of these
N=Z systems, we obtain the corresponding reduced density matrices
and implement our truncation.

The truncation is carried out to those $m$ states with the largest
density matrix eigenvalues. We do not worry about making sure that
at least one state is kept for each $N (or~Z)$ and $J$, as our
experience in {\em these} calculations indicates it is not
necessary.

\subsubsection{Sweeping}

As noted earlier, our procedure of separating neutron from proton
orbitals, rather than lining them up in a single chain, gives rise
to a three-block strategy in the sweeping process. We now expand
on what is meant by this, by reference to figure 3.

\begin{figure}
  \includegraphics[height=.2\textheight]{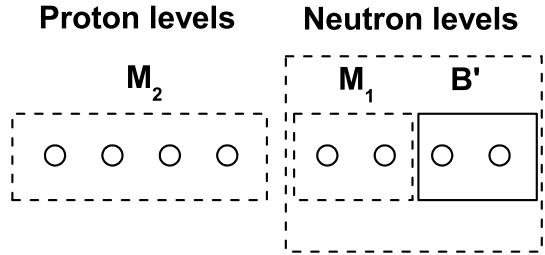}
  \caption{Schematic illustration of the three-block sweeping strategy used in the calculations reported herein.
  The enlarged neutron block $B'$ is immersed in a medium consisting of the remaining neutron levels
  (block $M_1$) and all the proton levels (block $M_2$).
}
\end{figure}

Let us assume that we are growing the system for one type of
particle (neutrons, for specificity), building from a block $B$ to
an enlarged block $B'$. The medium in which we carry out the
truncation of that enlarged block includes two components. One
component, denoted $M_1$ in the figure, involves the rest of the
neutron orbits, for which the corresponding optimal structure was
defined in the previous sweep (or in the warmup phase). The second
component, denoted $M_2$ in the figure, involves {\em all} of the
orbits of the protons, for which likewise the optimal structure
was obtained in the previous stage. Put another way, the states in
the superblock have the form

\[
\left| \left\{ \left[ \left( n_{\nu }\alpha _{\nu }J_{\nu }\right)
\left( 4-n_{\nu}~ \alpha'_{\nu}J'_{\nu}\right) \right]
^{J_{\pi}}\left(4 ~ \beta _{\pi}J_{\pi}\right) \right\}
_{M=0}^{L=0} \right\rangle ~,
\]
with $n_{\nu}$ neutrons in the enlarged neutron block $B'$,
$4-n_{\nu}$ neutrons in $M_1$ and $4$ protons in $M_2$.

By diagonalizing the hamiltonian in that basis, we arrive at a
ground state wave function
\begin{eqnarray}
&&|L=0~G.S. \rangle =\sum_{\alpha _{\nu }J_{\nu }
\alpha'_{\nu}J'_{\nu}\beta _{\pi}J_{\pi }K_{\nu}} X_{\alpha _{\nu
}J_{\nu }\alpha'_{\nu}J'_{\nu}\beta_{\pi}J_{\pi }} \nonumber\\
&&~\times \left| \left\{ \left[ \left( n_{\nu }\alpha _{\nu
}J_{\nu }\right) \left( 4-n_{\nu}~ \alpha'_{\nu}J'_{\nu}\right)
\right] ^{J_{\pi}}\left(4 ~ \beta _{\pi}J_{\pi}\right) \right\}
_{M=0}^{L=0} \right\rangle, \nonumber \\
\end{eqnarray}
from which the reduced density matrix associated with $B'$ can be
readily evaluated and then diagonalized for each $n_{\nu}$ and
$J_{\nu}$. Truncation to the $m$ states with the largest density
matrix eigenvalues gives the optimal structure of the truncated
block $B'$, which is stored for use in the next sweep stage.

\subsubsection{The whole process}

We now have in hand all that is needed to fully describe our
calculational algorithm. It involves the following steps.

\begin{enumerate}
\item Order the single-particle states. Indeed, in the results we
will report, we simply use the ordering $p_{1/2}$, $p_{3/2}$,
$f_{5/2}$, $f_{7/2}$ in terms of increasing angular momentum, with
no effort at optimization.

\item
For each $m$, we carry out the warmup as prescribed in Sec. IV A
storing the results ({\em i.e.,} the reduced matrix elements) for
each group of orbits.

\item
We then begin by sweeping downwards through the orbits, growing
from orbit $4$ to a block $B(4,3)$. The medium for implementing
this truncation involves the block $B(1,2)$ for the same type of
particle and the block $B(1,4)$ for the other, both of which were
obtained either in the warmup or in the previous sweep.

\item
We iterate the above step until all sets of orbits have been
treated.

\item
We then reverse and begin sweeping upwards, growing the block and
implementing a truncation based on its coupling to the other two
relevant blocks.

\item
We view a sweep down and the following sweep up as a single
sweeping stage. It involves six growing steps, each one giving a
ground state eigenvalue. We denote the lowest of these six
eigenvalues by $E_{min}^n$, where $n$ denotes the sweeping stage
number.

\item
As we systematically sweep up and down, we compare $E_{min}^n$
with $E_{min}^{n-1}$, {\em i.e,} we look at the change in the
minimum energy from one sweeping stage to the next. When this
change is sufficiently small, typically of order $10^{-4}~MeV$, we
stop the calculation. No more than 7 sweeping stages were required
to achieve this level of convergence in any of the calculations we
carried out.

\item
We then increment $m$ and repeat the set of steps $2 \rightarrow
7$.

\end{enumerate}

\section{Results}

Our results for the ground state are presented in table 1. The
exact calculation produces a ground state energy of $-32.95~MeV$.
The DMRG calculation with $m=50$ produces a result of
$-32.80~MeV$, about $150~keV$ from the exact result. In this case,
the largest matrix that had to be diagonalized had a dimension of
3,657. The conclusion is that the three-block JDMRG algorithm
gives a good reproduction of the exact results for the ground
state with a fairly small number of states retained in each block,
but by calculating matrices with a substantial fraction (roughly
8-9\%) of the full basis.

\begin{table}[h]
\caption{Results for the ground-state energy from the JDMRG
calculations described in the text. {\em Max~Dim} refers to the
maximum dimension of the hamiltonian matrix that required
diagonalization.} \label{tab:a} \vspace{0.1cm}
\begin{tabular}{|r|r|r|}
\hline
$m$ & $E_{GS}~(MeV)$ & $Max~Dim$ \\
\hline
20 & -32.28 & 1,591 \\
25 & -32.47 & 1,893 \\
30 & -32.57 & 2,327 \\
35 & -32.66 & 2,685 \\
40 & -32.72 & 3,109 \\
45 & -32.76 & 3,403 \\
50 & -32.80 & 3,657 \\
&&\\
$Exact$ & -32.95 & 41,355 \\

 \hline
\end{tabular}
\end{table}

For any value of $m$, we can calculate excitation energies as
well, by working in spaces that are based on the optimized blocks.
As a reminder, during the sweep procedure we diagonalize the
superblock hamiltonian in a basis in which all states of the added
level are included. Following the density-matrix-based truncation,
the enlarged block has fewer states, and often significantly
fewer. It is in this reduced space that we can recalculate the
ground state, by diagonalizing in the truncated $0^+$ space,  and
also carry out diagonalizations for other angular momenta. The
results for $m=50$ are illustrated in table 2, for the excitation
energies of the lowest $2^+$, $4^+$ and $6^+$ states.  The result
for the lowest $2^+$ state is in quite good agreement with the
exact result, despite the diagonalization of a matrix that is less
than 5\% of the full matrix. For the $4^+$ and $6^+$ states, the
results remain fairly good, although progressively worse than for
the lower $J$ values. These results can of course be improved by
further increasing $m$. Overall, we conclude that the method seems
capable of producing reasonable results for low-lying excited
states as well, even when we target the ground state only.

\begin{table}[h]
\caption{Results for the excitation energies of low-lying excited
states from the DMRG calculations described in the text at a value
of $m$=50. All results are in $MeV$.} \label{tab:b} \vspace{0.1cm}
\begin{tabular}{|r|r|r|}
\hline
$J^{\pi}$ & $JDMRG$ & $Exact$ \\
\hline
$2^+$ & 0.84 & 0.81 \\
$4^+$ & 2.02 & 1.82 \\
$6^+$ & 3.95 & 3.40 \\
\hline
\end{tabular}
\end{table}

\section{Summary and outlook}

In this paper, we have described an approach for solving the
nuclear shell-model problem in cases in which it cannot be treated
by exact diagonalization. The approach makes use of the Density
Matrix Renormalization Group to systematically truncate the space
in which diagonalizations are carried out on the basis of
well-defined dynamical considerations. The approach we develop
preserves rotational symmetry throughout, to avoid losses of
information associated with breaking of that critical nuclear
symmetry during the truncation process.

We test these ideas in the context of a large-scale, but exactly
diagonalizable, shell-model problem. Namely we apply it to the
well-deformed nucleus $^{48}Cr$ with a realistic KB3 hamiltonian.
For these test calculations, we used a three-block DMRG strategy,
in which neutron and proton levels were not mixed in blocks. This
leads to certain simplifications in the formalism, but at the cost
of having to treat larger superblock matrices.

The calculations are able to produce a fairly accurate
reproduction of the exact results for $^{48}Cr$, both for the
ground state and the first excited $2^+$ state, both members of
the ground state rotational band. It does this by retaining a
fairly small number of states in each block. However, because of
the three-block approach that was used, the dimensions of the
superblock matrices we needed to treat were still a sizable
fraction (5-10\%) of the full shell-model space. Results for the
lowest states of higher spins get progressively worse as the spin
increases.

Our work in the future will focus on two fronts. On the one hand,
we will continue to work with the three-block approach, addressing
such issues as ({\em i}) the possibility of improving convergence
by altering the order in which levels are included, building on
the work of Legeza and collaborators \cite{Legeza1}, and ({\em
ii}) the possibility of speeding up the calculations by replacing
$m$ as the convergence variable by the sum of neglected density
matrix eigenvalues \cite{Legeza2}. At the same time, we are now in
the process of developing a two-block JDMRG code \cite{Thakur}, in
which blocks containing both neutron and proton orbits are built.
This should enable us to dramatically increase the number of
states retained in a block for a given size of the full superblock
matrix. This will be critical as we consider the application of
these methods to heavier nuclei, the ultimate goal of the project.

\begin{flushleft}
{\bf Acknowledgements}
\end{flushleft}
This work was supported in part by the US National Science
Foundation under grant \# PHY-0140036. We acknowldege with deep
appreciation the many contributions of Jorge Dukelsky to this
project. We also thank Alfredo Poves for providing the matrix
elements and exact results used in this work and Larisa Pacearescu
for her contributions in the early stages. Finally, we are
grateful to Ian McCulloch for valuable discussions on the
non-Abelian DMRG method.

\end{document}